\documentclass[10pt,a4paper]{iopart}
\usepackage[left=2.9cm,right=2.8cm,top=2cm,bottom=1.5cm]{geometry}
\usepackage{graphicx}
\usepackage{caption}
\usepackage[colorlinks]{hyperref}
\hypersetup{colorlinks,breaklinks,
            citecolor=[RGB]{25,132,185},
            urlcolor=[RGB]{25,132,185},
            linkcolor=[RGB]{25,132,185}
            }

\usepackage{cite}

\begin{document}

\title[]{Vortex detection in atomic Bose--Einstein condensates using neural networks trained on synthetic images}

\author{Myeonghyeon Kim,$^{1,2}$ Junhwan Kwon,$^{1}$ Tenzin Rabga,$^{2}$ and Y. Shin$^{1,2,3,*}$}
\address{$^1$Department of Physics and Astronomy, Seoul National University, Seoul 08826, Korea}
\address{$^2$Center for Correlated Electron Systems, Institute for Basic Science, Seoul 08826, Korea}
\address{$^3$Institute of Applied Physics, Seoul National University, Seoul 08826, Korea}
\ead{\href{yishin@snu.ac.kr}{yishin@snu.ac.kr}}



\begin{abstract}
Quantum vortices in atomic Bose--Einstein condensates (BECs) are topological defects characterized by quantized circulation of particles around them. In experimental studies, vortices are commonly detected by time-of-flight imaging, where their density-depleted cores are enlarged. In this work, we describe a machine learning-based method for detecting vortices in experimental BEC images, particularly focusing on turbulent condensates containing irregularly distributed vortices. Our approach employs a convolutional neural network (CNN) trained solely on synthetic simulated images, eliminating the need for manual labeling of the vortex positions as ground truth. We find that the CNN achieves accurate vortex detection in real experimental images, thereby facilitating analysis of large experimental datasets without being constrained by specific experimental conditions. This novel approach represents a significant advancement in studying quantum vortex dynamics and streamlines the analysis process in the investigation of turbulent BECs.

\end{abstract}

\section{Introduction\label{Sec:introduction}}

Quantum vortices are topological defects in superfluid systems, characterized by phase singularities in the superfluid order parameter. These vortices exhibit quantized circulation of particles around them due to the integer phase winding, leading to a depleted particle density at their cores, reminiscent of the eye of a tornado. Quantum vortices play an important role in various superfluid transport phenomena, and extensive research has been conducted on their dynamic properties in different superfluid systems such as liquid helium~\cite{donnelly1991}, superconductors~\cite{degnnes1966}, and atomic Bose--Einstein condensates (BECs)~\cite{matthews1999}. Atomic BECs, in particular, offer several advantages in the study of quantum vortex dynamics, as they can be effectively described by the Gross--Pitaevskii equation (GPE) within the mean-field approximation, allowing direct comparisons between theoretical predictions and experimental observations~\cite{fetter2001}. Additionally, resonant imaging techniques allow for direct visualization of quantum vortices in atomic BECs during a time-of-flight (ToF) expansion, where the vortex core size becomes larger than the imaging device's resolution~\cite{lundh1998,dalfovo2000,seo2014}. Due to these advantages, there has been active research on the properties of quantum vortices in atomic BECs\cite{fetter2001,srinivasan2006} as well as investigations of related phenomena such as rotating BECs~\cite{matthews1999,madison2000,abo-shaeer2001,hodby2001,engels2002}, vortex shedding~\cite{inouye2001,neely2010,kwon2015,kwon2016,lim2022}, 2D superfluidity~\cite{hadzibabic2006,schweikhard2007,choi2013,seo2017}, turbulence~\cite{tsatsos2016,henn2009,neely2013,kwon2014,johnstone2019,gauthier2019}, and spontaneous defect formation~\cite{weiler2008,ko2019,goo2021,goo2022}. These studies have significantly advanced our understanding of the superfluid physics associated with these intriguing topological defects.

When imaging vortices in a BEC, the vortex cores must be identified through appropriate image analysis. One simple method is manual counting by a human observer; however, this can be exceedingly time-consuming when dealing with large datasets, especially in BEC samples containing numerous vortices. To address this issue, previous studies have developed several image processing algorithms for automated detection of density-depleted cores. In rotating BECs, because vortices form a lattice structure with distinct separation, Gaussian and Laplacian filters can readily locate local density minima, which coincide with single vortex cores~\cite{rakonjac2016}. However, in turbulent BECs, vortex distribution is irregular, and vortices may not be well-separated, leading to multiple vortices within a single density-depleted region. In such cases, the number of vortices can be estimated by measuring the area of the density-depleted regions from a binarized image of the density using a given threshold~\cite{kwon2016,ko2019}. However, accurately determining the locations of vortex cores remains a challenge even with this method.

Recent advances in artificial intelligence and neural network technology have enabled the use of machine learning algorithms in cold atom research. These algorithms have been used to optimize experimental parameters~\cite{wigley2016,barker2020,ness2020,davletov2020} and analyze experimental and numerical data~\cite{rem2019,bohrdt2019,kaming2021,hofer2021,guo2021,lode2021,sharma2022,guo2022,metz2021}, providing high accuracy and efficiency. Metz \textit{et al.}~\cite{metz2021} recently demonstrated the successful application of a convolutional neural network (CNN) to detect vortices in numerically generated BEC images. This successful debut of the CNN algorithm in numerical studies of BECs encourages its use in real experiments. However, several challenges must be addressed when using the CNN algorithm to analyze experimental data. The actual experimental images have a lower resolution than the numerical images because of the limitations of the experimental setup. Additionally, during ToF expansion before imaging, BEC samples undergo changes that can lead to blurring of the vortex core density gradient and, in some cases, the production of density ripples around the core~\cite{seo2014}. 

Another significant challenge is to obtain a sufficiently large and labeled dataset to train CNN. Machine learning algorithms often require a substantial amount of labeled data for effective training, but collecting such data in real-world experiments can be time-consuming and costly, which is particularly relevant in atomic BEC experiments. Recently, to tackle the issue of obtaining a costly labeled training set for machine learning algorithms, the use of synthetic data has been proposed~\cite{nikolenko2021,demelo2022,lu2023}. This is especially attractive as the GPE can generate synthetic BEC images for CNN training with parameters that are similar to those of the experimental setup. Moreover, an automated algorithm can be used to accurately identify the ground truth of synthetic images, i.e. vortex positions, eliminating the need for manual labeling and providing efficiency and reduced bias.

In this paper, we present our work on the use of CNNs trained on synthetic images to detect vortices in experimental BEC images. To address the issue of low resolution in experimental images, we employ an upsampling process and evaluate the CNN's performance by varying the upsampling size. Additionally, we optimize the CNN's performance in recognizing vortices in the experimental BEC images by adjusting image variables in the training dataset, such as the noise strength and the criterion for defining the BEC boundary. Our results demonstrate that, with proper parameter tuning, CNNs trained on synthetic images can accurately and effectively detect vortices in experimental BEC images, providing a useful tool for analyzing quantum vortices in real experimental data.

The remainder of this paper is organized as follows. Section \ref{Sec:methods} provides a detailed description of the experimental setup used to generate BECs with vortices and the GPE simulation process to create synthetic BEC images that closely resemble the experimental conditions. Additionally, we explain the CNN training procedure using these simulated images. In Section \ref{Sec:results}, we present the results of applying CNN to detect vortices in experimental images. We analyze the effect of tuning parameters on CNN's performance and discuss how these parameters affect the accuracy of vortex detection in the experimental data. Finally, in Section \ref{Sec:conclusion}, we summarize our findings and provide an outlook on the use of machine learning algorithms for vortex detection in the study of vortex dynamics in atomic BECs.

\section{Methods\label{Sec:methods}}

\subsection{Experimental image data\label{Subsec:experiments}}

We experimentally produce BECs with multiple vortices by rapidly cooling an atomic thermal gas, causing it to undergo a phase transition into the superfluid phase. During the quench process, the superfluid order parameter of the system develops independently in local areas, leading to the emergence of topological defects at the interfaces between these phase domains. This phenomenon is called the Kibble--Zurek mechanism~\cite{kibble1976,zurek1985}. In the case of a superfluid system, these topological defects appear as quantum vortices. 

The experiment begins with the preparation of a thermal gas of $^{87}$Rb atoms in an optical dipole trap (ODT) with a highly oblate and elongated geometry, as described in~\cite{lim2021}. The trapping potential of the ODT is expressed as
\begin{equation}
    V(x,y,z)=\frac{1}{2}m\omega^2 R_x^2 \Bigg[ \left ( \frac{x}{R_x} \right )^2  +\left( \frac{y}{R_y}\right)^{3.9} + \left ( \frac{z}{R_z} \right )^2 \Bigg],
\end{equation}
where $\omega \approx 2\pi \times 7$~Hz is the trapping frequency for the $x$ direction, $m$ is the mass of a $^{87}$Rb atom, and $\{R_x, R_y, R_z\} \approx \{65, 244, 2.8\}\,{\rm \mu m}$ are the Thomas--Fermi radii of the final condensate. The ODT provides strong confinement along the $z$ direction and this highly oblate geometry causes the vortex lines to be energetically aligned along the tightly confining $z$-axis. 

We rapidly cool the thermal gas by decreasing the trap depth of the ODT from $1.15 U_c$ to $0.27 U_c$, where $U_c$ is the critical trap depth at which the sample starts to undergo a phase transition. During the quench, the temperature of the sample is regulated by the depth of the trap through evaporation. To facilitate the formation of quantum vortices, we keep the sample in the ODT for an additional 1.25~s after the quench. We then release the trapping potential to allow the BEC to expand. During this expansion, the vortex cores expand faster, becoming visible in the subsequent imaging. After a 40-ms ToF expansion, we perform absorption imaging of the BEC along the $z$-axis to capture the density distribution of the BEC, which allows us to detect the presence and spatial arrangement of quantum vortices. The cooling speed is controlled by the decrease time of the trap depth, which ranges from 0.8~s to 11~s, resulting in a variable vortex number in the BEC. For our fastest quench, the vortex number exceeds 60~\cite{goo2021,goo2022}. The typical number of atoms in the condensate is approximately $N=9.6\times 10^6$.

\begin{figure*}[t]
	\includegraphics[width=146mm]{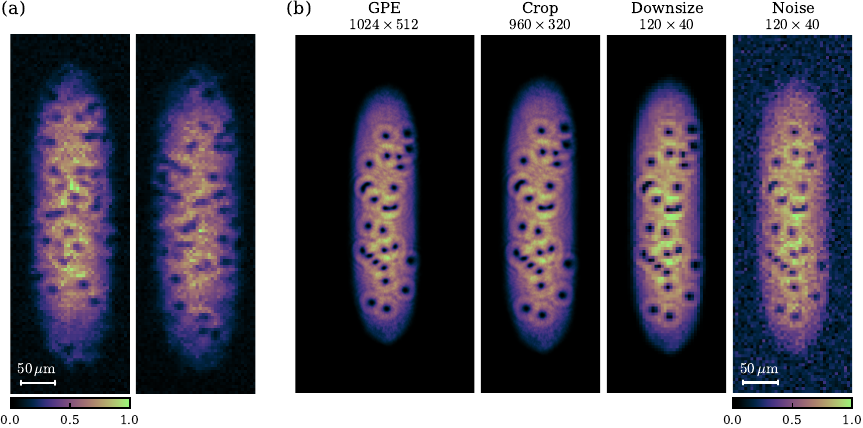}
	\caption{Images of atomic Bose--Einstein condensates (BECs) containing quantum vortices. (a) Representative experimental images: A thermal Bose gas was rapidly cooled to the superfluid phase, leading to the formation of a BEC with multiple vortices. The BEC images were acquired after a 40-ms time-of-flight (ToF) expansion, which highlights the expanded, density-depleted cores of the vortices. The sample images have been cropped to a size of 120$\times$40 pixels for analysis. (b) Synthetic image generation: Starting from a high-resolution BEC image obtained through numerical solution of the Gross--Pitaevskii equation (GPE), subsequent image processing steps are shown from left to right. The high-resolution image is cropped and downsized to match the resolution and size of the experimental images, and Gaussian noise is then added to each pixel of the image to simulate realistic experimental conditions.}
    \label{Fig:figure1}
\end{figure*}

In Figure~\ref{Fig:figure1}(a), we display examples of experimental images in which the vortex cores are identified as localized areas of reduced density. The images are presented as 2D optical depth (OD) arrays and have been cropped to a size of 120$\times$40 pixels, with each pixel being approximately 4.3~$\mu$m in size. This cropping ensures that the BEC sample is centered and enclosed within the image. To remove any outlier pixels that may have been caused by noise during the calculation of OD, we trim values that differ by more than four standard deviations from the surrounding values, replacing them with the average value of neighboring pixels. We then normalize the images so that the minimum pixel value is set to zero and the maximum is set to one, creating a consistent scale for analysis.

\subsection{Generating synthetic training sets}

The dynamics of weakly interacting atomic BECs can be effectively described by the GPE~\cite{fetter2001}. This equation takes the form
\begin{equation}
i\hbar \frac{\partial}{\partial t} \psi(x,y,z,t) = \Bigg[ -\frac{\hbar^2}{2m}\nabla^2 + V(x,y,z) + \frac{4\pi \hbar^2 N a_s}{m} \psi ^2 \Bigg] \psi(x,y,z,t),
\end{equation}
where $\psi$ is the condensate wave function and $a_s$ is the scattering length of the atoms. The BEC dynamics in the $z$ direction is highly suppressed due to tight confinement, so we simplify the equation by integrating out the $z$-dependent part of the wave function, resulting in an effective equation for a 2D complex wave function $\psi(x,y)$. The numerical solution of the 2D GPE is obtained on a grid space of 1024$\times$512 with a cell size of 0.54~$\mu$m, which has a much higher resolution than our experimental images. The healing length of the condensate in the experiment was 0.26~$\mu$m and such high resolution is necessary for the GPE simulation. After determining the complex wave function of the system from the GPE, we calculate the density profile of the BEC by taking the squared norm of the wave function $|\psi(x,y)|^2$.

To create synthetic images of BECs with quantum vortices, we first calculate the ground state of the system using the imaginary-time evolution of the GPE~\cite{muruganandam2009,vudragovic2012}. Then, we imprint phase singularities onto the condensate wave function. The phase singularity mask is given by $\theta(x,y)=\sum_{i=1}^{N_v} s_i \arctan(x-x_i,y-y_i)$, where $(x_i,y_i)$ is the position of the $i$-th phase singularity and $s_i=\pm 1$ is the sign of phase winding randomly chosen. The number of singularities, $N_v$, imprinted on the condensate ranges from 0 to 100, and their locations are randomly distributed throughout the sample area. After phase imprinting, we evolve the condensate with the new phase profile for a short hold time using the real-time evolution of the GPE, during which phase singularities develop into vortices in the system. Then, to account for the effects of the time-of-flight (ToF) imaging process, where the density profile around the vortex cores are slightly altered, we allow the sample to expand by suddenly releasing the trapping potential. In the free expansion, we gradually reduce the interaction strength for the same ToF time to simulate the effect due to the fast expansion along the $z$ direction~\cite{lundh1998, dalfovo2000}. Finally, we create a density image through the squared norm of the wavefunction. Figure~\ref{Fig:figure1}(b) shows typical simulated BEC images.

\begin{figure*}[t]
	\includegraphics[width=146mm]{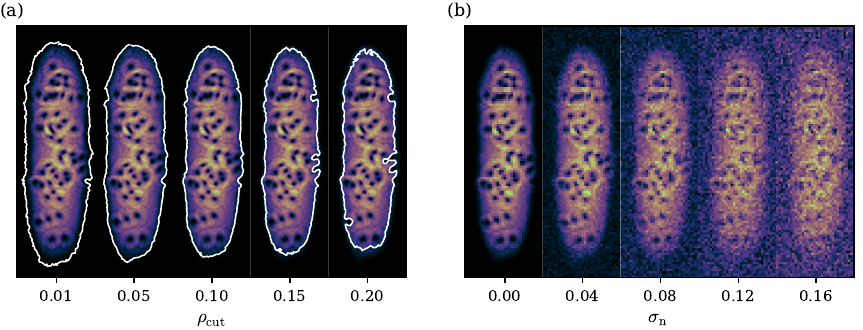}
	\caption{Tuning parameters for the training set of synthetic images. (a) Determining the effective sample area: The region where the BEC density exceeds a specified cutoff density $\rho_{\rm cut}$ is considered as the effective sample area. Images illustrate different boundaries of the BEC for various density cutoff values $\rho_{\rm cut}$. Each image was normalized to have a maximum value of unity. Only phase singularities within this defined area are labeled as true vortices. (b) Introducing noise: To replicate realistic experimental conditions, Gaussian random values with a standard deviation $\sigma_{\rm n}$ were added to each pixel of the normalized simulated images. Images with different noise levels are shown. The images were normalized again after adding the noise.}
    \label{Fig:figure2}
\end{figure*}

After obtaining a synthetic BEC image, we label it with the positions of the vortices in the simulated BEC sample. We know the initial positions of the phase singularities on the condensate, but the vortices may have shifted and dissipated during the evolution after the phase-imprinting sequence. To identify the vortex positions, we use a brute-force algorithm based on the phase information of the wave function~\cite{metz2021}. The algorithm searches for all local density minima as potential vortices and calculates the integral of the phase gradient $\nabla \left({\rm arg}(\psi) \right)$ along the circumference of each candidate point. If the integral is non-zero, the point is identified as a vortex. To ensure that only the vortices within the BEC are accurately labeled, we define a sample area criterion using a cutoff density, denoted by $\rho_{\rm cut}$, and exclude any singularity in the outer region of the sample with density $\rho \leq \rho_{\rm cut}$ from the labeled vortex list. Figure~\ref{Fig:figure2}(a) shows the change of the sample area enclosed by the contour ($\rho=\rho_{\rm cut}$) as the value of $\rho_{\rm cut}$ is adjusted. Here, the sample density is normalized so that the maximum density, ${\rm max}( |\psi(x,y)|^2)$, is set to one.

The last step in creating a synthetic training set is to pre-process the simulated images to make them look like the ones obtained in the experiment. Initially, the simulated images are cropped to 960$\times$320 pixels and then downsized to 120$\times$40 pixels, so that they have the same resolution and size as the experimental images. After that, Gaussian noise is added to each pixel of the image with a uniform noise strength across the image. The Gaussian noise has a zero mean and a standard deviation denoted as $\sigma_{\rm n}$. By adjusting the tuning parameter $\sigma_{\rm n}$, the noise level of the images can be changed, as shown in Figure~\ref{Fig:figure2}(b). Finally, the image is normalized to set the minimum and maximum pixel values to zero and one, respectively, completing the pre-processing sequence (Figure~\ref{Fig:figure1}(b)).

\subsection{CNN training and evaluation}

Our vortex detection algorithm is based on the CNN model proposed by Metz \textit{et al.}~\cite{metz2021}. This CNN architecture consists of seven convolutional layers and three maxpool layers. The stride of the layers is 1, except for the first two maxpool layers, which have a stride of 2 to reduce the output size. The CNN takes greyscale density images as a single input channel, and it has three output channels, each reduced both in height and width by a factor of four compared to the input channel. Each 4$\times$4 grid cell of the input, serving as the unit window for vortex detection, is coarsened into a set of three values for the output channels, which represent the probability of a vortex core being in the grid cell and the rescaled $x$ and $y$ positions of the vortex core within the grid cell, respectively. When the size of the input image is $\rm H \times W$, the resulting output tensor has dimensions of $\rm H/4 \times W/4 \times 3$, which are denoted as $\widetilde{Y}_{ijk}$, where $i$ and $j$ are the position indices of a 4$\times$4 grid cell, and $k$ is the index of the output channels. For ground-truth data of the vortex positions, the array $Y_{ijk}$ is created, where $Y_{ij1}=1$ when a grid cell contains a vortex and $Y_{ij1}=0$ when a grid cell contains no vortex. The loss function used for training the CNN is defined as follows: 
\begin{eqnarray}
L = \sum_{\rm batch}\sum_{i,j} \Bigg[ &-w_{1}Y_{ij1}\log(\widetilde{Y}_{ij1}) -(1-Y_{ij1})\log(1-\widetilde{Y}_{ij1}) \\ \nonumber &+w_{2}Y_{ij1} \bigg( {(Y_{ij2}-\widetilde{Y}_{ij2})}^2+{(Y_{ij3}-\widetilde{Y}_{ij3})}^2 \bigg) \Bigg], \end{eqnarray}
where $w_1$ and $w_2$ are hyperparameters that control the weighting of each term. In this study, $w_1$ and $w_2$ are set to 10, as in~\cite{metz2021}.

We apply upsampling to the experimental and synthetic images to feed them to the CNN. Nearest-neighbor upsampling is used, where each pixel in the original image is repeated $n_{\rm up}$ times in both the row and column directions without any interpolation. The original size of the images is 120$\times$40 pixels, and after upsampling, they become $120n_{\rm up}\times40n_{\rm up}$ in size. The value of $n_{\rm up}$ can be adjusted to change the resolution of the images, but it should be noted that the images with different $n_{\rm up}$ values look the same because no interpolation is used during the upsampling process.

The CNN is trained on a dataset of 2000 synthetic images, with a batch size of 100. The training process is conducted for 300 epochs, with an initial learning rate of $\eta = 10^{-3}$. After 300 epochs, the learning rate is decreased to $10^{-4}$. To ensure a fair comparison of the performance of CNNs trained with various parameter combinations, the early stopping method is used with a patience value of 50 epochs, without fixing the training epoch. This means that, while monitoring the loss function, if it does not show further reduction over the course of 50 epochs, the algorithm regards it as having reached its minimum and concludes the training process.

The performance of the CNN is evaluated by comparing the number of vortices identified by the machine to the number counted by a human observer. To quantify the discrepancy between the machine and human counts, we calculate the root mean squared error (RMSE) value that is given by 
\begin{equation}
{\rm RMSE} = \left( \frac{\sum_{i}^{n}{\left[ N_{\rm m}(i)-N_{\rm h}(i) \right]^2}}{n} \right)^{1/2},
\end{equation}
where $N_{\rm m}(i)$ is the machine count for the \textit{i}-th image, $N_{\rm h}(i)$ is the corresponding human count, and $n=350$ is the number of experimental images used for the test. Out of the 350 experimental images, the distribution of $N_{\rm h}$ is as follows: 56.2\% have 0 to 20 vortices, 18.9\% have 20 to 40 vortices, 20.6\% have 40 to 60 vortices, and 4.3\% have 60 to 80 vortices. The RMSE value provides an estimate of the average difference between the machine and human counts per image.  
The confidence threshold is set to 0.5 for the CNN's vortex detection.

\section{Results and Discussion\label{Sec:results}}

\subsection{Parameter tuning of synthetic training sets}

\begin{figure*}[t]
	\includegraphics[width=146mm]{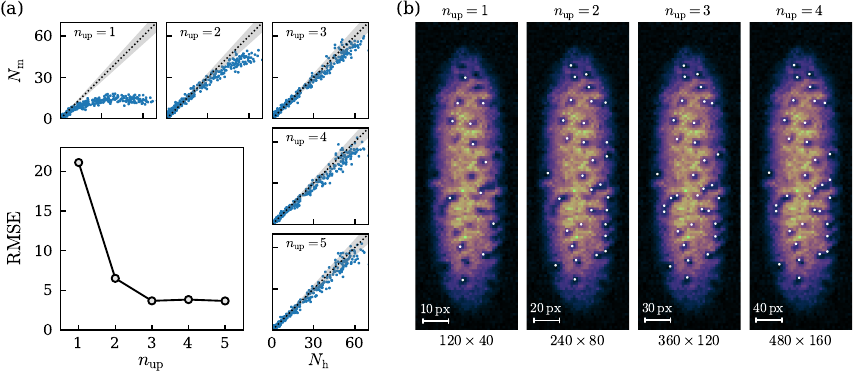}
	\caption{Performance of the CNN with varying upsampling size $n_{\rm up}$. (a) Root mean squared error (RMSE) values for different $n_{\rm up}$ settings. $\sigma_{\rm n}=0.07$ and $\rho_{\rm cut}=0.01$. The scatter plots in the upper and right sides compare the vortex number count $N_{\rm m}$ obtained using the CNN and the vortex number count $N_{\rm h}$ determined by a human observer, for different $n_{\rm up}$ values. The dotted lines and gray regions in the plots indicate the $N_{\rm m}=N_{\rm h}$ line and $\pm 10\%$ of $N_{\rm h}$, respectively. (b) Detection results for an experimental image using CNNs trained on images upsampled at different $n_{\rm up}$ sizes. Detected vortices are indicated by white dots. 
    }
    \label{Fig:figure3}
\end{figure*}

\begin{figure*}[t]
	\includegraphics[width=146mm]{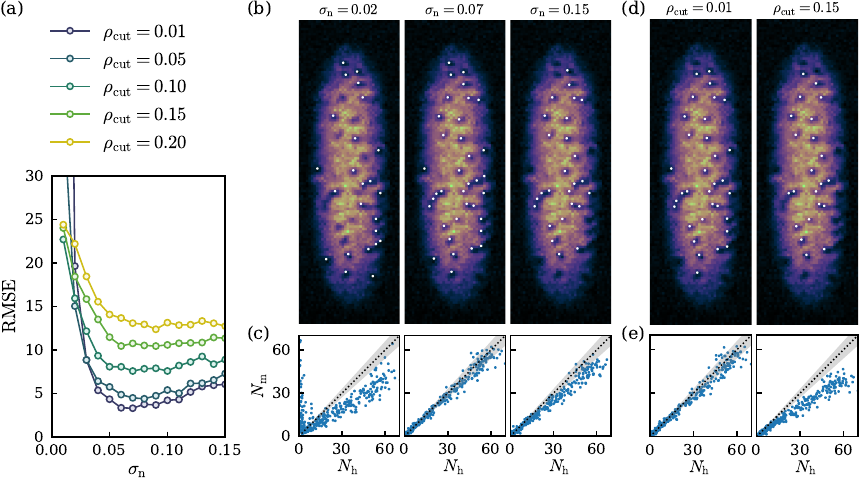}
	\caption{Performance of the CNN with varying noise strength $\sigma_{\rm n}$ and cutoff density $\rho_{\rm cut}$. (a) RMSE values as a function of $\sigma_{\rm n}$ for different $\rho_{\rm cut}$ values. The RMSE has a minimum value of $\approx 3$ at $\sigma_{\rm n}=0.07$ and $\rho_{\rm cut}=0.01$, suggesting an optimal combination of noise strength and density cutoff for accurate vortex detection. (b) CNN detection results for $\sigma_{\rm n}= 0.02, 0.07, 0.15$ with fixed $\rho_{\rm cut}=0.01$, and (c) the corresponding scatter plots of $N_{\rm h}$ and $N_{\rm m}$. (d) CNN detection results for $\rho_{\rm cut}=0.01, 0.15$ with fixed $\sigma_{\rm n}=0.08$, and (e) the corresponding scatter plots. The CNN trained with images with high $\rho_{\rm cut}$ tends to miss vortices at the boundary region of the BEC sample. The dotted lines and the gray regions in (c) and (e) indicate the $N_{\rm m}=N_{\rm h}$ line and $\pm 10\%$ of $N_{\rm h}$, respectively, same as in Figure~\ref{Fig:figure3}(a). The upsampling size was fixed at $n_{\rm up}=4$.}
    \label{Fig:figure4}
\end{figure*}

The performance of the machine for vortex detection is affected by the tuning parameters such as the upsampling size $n_{\rm up}$ for imaging resolution, the cutoff density $\rho_{\rm cut}$, and the noise strength $\sigma_{\rm n}$ in the formation of synthetic training sets. In this section, we explain how the machine performance is changed by the three parameters. We train various CNNs with different parameter settings and compare their RMSE values to determine the optimal parameter combination for accurate vortex detection.

We first investigate the effect of image resolution on machine performance by adjusting the upsampling size $n_{\rm up}$ from one to five, while keeping $\sigma_{\rm n}=0.07$ and $\rho_{\rm cut}=0.01$ fixed. The results of the evaluation are shown in Figure~\ref{Fig:figure3}. As the size of the upsampling increases from $n_{\rm up}=1$ to 2, the machine performance improves significantly, reducing the RMSE value by a factor of about 3. Further increases in $n_{\rm up}$ result in gradual improvements in detection performance, and for $n_{\rm up}>3$, the RMSE value converges to approximately 3. The scatter plots of Figure~\ref{Fig:figure3}(a) compare the vortex number counts $N_{\rm m}$ and $N_{\rm h}$ for various $n_{\rm up}$ to show the details of the machine performance. When $n_{\rm up}=1$, i.e., without upsampling, the machine exhibits poor performance, particularly for cases with a high vortex number, indicating that the image resolution is not sufficient to resolve vortices when clustered with high density (Figure~\ref{Fig:figure3}(b)). It is interesting to note that since the 4$\times$4 grid cell is used as the unit window for single vortex detection in the CNN model, $n_{\rm up}=4$ corresponds to having a resolution of 1 pixel in the bare experimental image. This seems to explain the RMSE convergence for $n_{\rm up} \geq 3$. The results for various $n_{\rm up}$ suggest that we can utilize the CNN detection algorithm effectively for low-resolution experimental images by simple upsampling without interpolation. The required upsampling size may be smaller in experimental settings with better resolution.

Next, we explore the effect of the parameters $\sigma_{\rm n}$ and $\rho_{\rm cut}$ on machine performance. We measure the RMSE value for a range of noise strength $\sigma_{\rm n}$ from 0.01 to 0.15, for different $\rho_{\rm cut}$ values and a fixed $n_{\rm up}=4$. The results shown in Figure~\ref{Fig:figure4}(a), demonstrate a U-shaped dependence on the noise strength $\sigma_{\rm n}$, with an optimal noise strength of around 0.07, regardless of the density cutoff value used. At $\sigma_{\rm n}=0.07$ and $\rho_{\rm cut}=0.01$, the RMSE has a minimum value of approximately 3. Interestingly, a CNN trained with low-noise images tends to underestimate the number of vortices, although it often overestimates for images with a few vortices (Figure~\ref{Fig:figure4}(c)). This behavior appears counterintuitive because it is expected that, trained with low-noise images, the machine might interpret strong noises as vortices, leading to overcounting. When the machine is trained with high-noise images, it shows an underestimate of the vortex number for the entire experimental image test set, reflecting its low sensitivity to vortices, ignoring them as noises (Figures~\ref{Fig:figure4}(b) and \ref{Fig:figure4}(c)).

The sample-area criterion, represented by the cutoff density $\rho_{\rm cut}$ in the synthetic images, also has a significant impact on the CNN performance. The density cutoff $\rho_{\rm cut}$ determines how close to the sample boundary is a vortex labeled in the synthetic image data. For noise strengths of 0.03 or higher, the CNN's performance improves as the cutoff density cutoff decreases, even to 0.01 (Figure~\ref{Fig:figure4}(a)). As shown in Figure~\ref{Fig:figure4}(d), the CNN trained for higher $\rho_{\rm cut}$ tends not to count density dips near the sample boundary as vortices. It is worth noting that Metz \textit{et al.}~\cite{metz2021} found that a CNN works efficiently with $\rho_{\rm cut}=0.15$ to analyze simulated BEC images. The optimal value of the cutoff density may differ under various task conditions.

\subsection{Noise strength mixing method}

The peculiar dependence of the machine performance on noise strength $\sigma_{\rm n}$ observed in the previous section motivates further investigation to find a more effective implementation of signal noise in the synthetic training dataset. Our current approach to creating a synthetic dataset involves adding Gaussian noise with a uniform strength across the entire image. However, it should be noted that noise in actual experimental images is the result of a complex combination of multiple factors, and its strength can vary spatially over the sample and fluctuate for different realizations. The primary source is atom-shot noise, which is caused by the quantized nature of atoms and leads to fluctuations in atom density. In the simplest scenario of a non-interacting ideal gas, the counting of atoms, $N$, within a specific volume follows a Poisson distribution, resulting in a variance of atom number $(\Delta N)^2 \sim \bar{N}$ with $\bar{N}$ being the mean value of $N$. Even when interactions among particles are taken into account, atom-shot noise remains a significant source of noise for atom image data~\cite{esteve2006,sanner2010}. Additionally, the inherent Poisson statistics associated with photon counting in the imaging process introduce another unavoidable source of noise as well as other technical noises.

\begin{figure*}[b]
	\includegraphics[width=146mm]{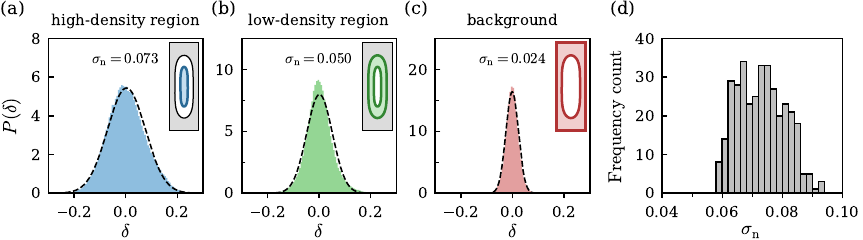}
	\caption{Characteristics of noise in experimental images. (a-c) Normalized noise distributions in three regions of an image: (a) center region with high atom density, (b) sample boundary region with low atom density, and (c) background region without atoms. Noise $\delta$ was measured as the difference between the pixel's own value and the average value of its 4 nearest and 4 next-nearest neighboring pixel values. The noise strength $\sigma_{\rm n}$ was estimated by calculating the standard deviation of $\delta$. Dashed lines represent Gaussian curves with standard deviations set to $\sigma_{\rm n}$. The insets indicate the corresponding regions in the image. (d) Frequency counts of $\sigma_{\rm n}$ in the high-density center region for the 350 experimental images.}
    \label{Fig:figure5}
\end{figure*}

To gain insight into the noise characteristics of our experimental images, we measure the noise strength and its spatial variations over the image. For each pixel, we calculate the difference $\delta$, between its value and the average value of its 4 nearest and 4 next-nearest neighboring pixels. We then estimate the noise strength $\sigma_{\rm n}$ in a region of interest by taking the standard deviation of $\delta$ in the region. Figures~\ref{Fig:figure5}(a), \ref{Fig:figure5}(b), and \ref{Fig:figure5}(c) show the probability distributions of $\delta$ obtained from three different regions in the image; a high-atom-density center region, a low-atom-density sample boundary region, and a background region without atoms, respectively. The noise distributions are found to be well modeled by Gaussian profiles, and the measured $\sigma_{\rm n}$ values increase with increasing atom density, which is in agreement with the predictions from atom-shot and photon-shot noises. In particular, the noise strength in the high-density center region is $\sigma_{\rm n}=0.073$, close to the optimal noise strength of 0.07 for synthetic images, which explains the optimization. Additionally, Figure~\ref{Fig:figure5}(d) displays the histogram of the noise strength in the high-density region across all experimental images, showing that the noise strength fluctuates from one image to another, centered around 0.073.

\begin{figure*}[t]
	\includegraphics[width=146mm]{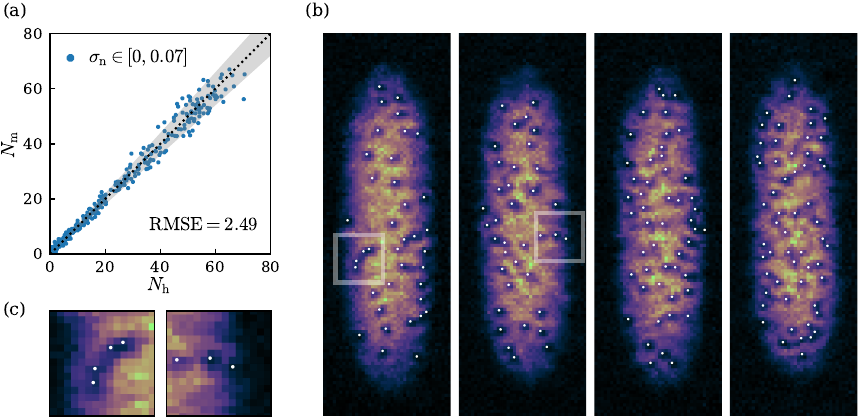}
	\caption{Performance of the CNN trained using a noise strength mixing method. (a) Scatter plot between $N_{\rm h}$ and $N_{\rm m}$, where CNN was trained on synthetic images with a range of $\sigma_{\rm n}$ values from 0.00 to 0.07 and a fixed $\rho_{\rm cut}=0.01$. The dotted line and the gray region indicate the $N_{\rm m}=N_{\rm h}$ line and $\pm 10\%$ of $N_{\rm h}$, respectively. (b) CNN detection results for four experimental image examples. (c) Examples of complex structures in turbulent BECs. The regions corresponding to these examples are marked by white boxes in (b). The vortices detected by CNN are indicated by white dots.}
    \label{Fig:figure6}
\end{figure*}

The proposition that the best training dataset is one collected from real task situations suggests that it is desirable to create a synthetic image dataset that includes all the noise characteristics of experimental images. However, it is difficult to incorporate every noise factor into synthetic images through a comprehensive theoretical analysis. As an alternative, we construct a training set with synthetic images that have different noise strengths while maintaining uniform Gaussian noise. We train the CNN using a set of 2000 training images with varying noise levels, ranging from 0.00 to 0.07, where $n_{\rm up}=4$ and $\rho_{\rm cut}=0.01$. The resulting RMSE value is 2.49, which is approximately 15\% lower than the minimum RMSE obtained when using the optimally selected uniform noise strength $\sigma_{\rm n}=0.07$ (Figure~\ref{Fig:figure6}(a)). This improvement indicates that the non-uniform noise strength of the experimental images can be better addressed by the noise strength mixing method.

Turbulent BECs contain a large number of vortices distributed in a disorganized manner (Figure~\ref{Fig:figure6}(b)) and they lead to experimental images with intricate patterns, such as linear chains of density dips created by multiple vortices and faint cracks at the sample boundaries that resemble vortices (Figure~\ref{Fig:figure6}(c)). This complexity makes it hard to accurately count the vortices, even for a human observer. Nevertheless, our vortex detection algorithm performs well, with an RMSE of $\approx 3$, which is more than satisfactory for the purposes of this study.

\section{Summary and Outlook\label{Sec:conclusion}}

We have presented an effective machine learning-based vortex detection algorithm for experimental images of atomic BECs. By training a CNN with synthetic images generated from the GPE, we achieved accurate and efficient vortex detection. We explored the effects of three key parameters in the synthetic training set, namely, the upsampling size, noise strength, and cutoff density, to optimize the CNN performance. We found that the low-resolution of experimental images can be overcome by the upsampling process, while a diverse training set with various noise strengths improved the algorithm's ability to distinguish real vortices from noise-induced artifacts. Additionally, adjusting the cutoff density enabled us to accurately label the vortices within the sample area, thereby refining the CNN's performance near the sample boundary.

Our vortex detection method has a number of advantages for the experimental study of vortex dynamics in atomic BECs. CNN's capability to accurately detect vortices in low-resolution experimental images, combined with its high speed, makes it possible to quickly process a large amount of image data, which could be beneficial for statistical investigations and correlation studies on turbulent BECs~\cite{gomez2020,liu2021,kim2022,delcampo2022,rabga2023}. Furthermore, the synthetic training dataset can be easily adjusted to different experimental and imaging configurations, allowing for broad and rapid applications of the algorithm to various experimental studies. Finally, we anticipate that the algorithm can be extended to analyze time-resolved data and track vortex trajectories in dynamic BEC systems. The combination of accuracy, efficiency, and flexibility makes the algorithm a valuable tool for future studies of vortices and, possibly, other topological defects in BEC systems.

\ack
This work was supported by the National Research Foundation of Korea (Grant No. NRF-2023R1A2C3006565) and the Institute for Basic Science in Korea (Grant No. IBS-R009-D1).

\section*{References}
\bibliography{ref.bib}

\providecommand{\newblock}{}
\begin{thebibliography}{10}
\expandafter\ifx\csname url\endcsname\relax
  \def\url#1{{\tt #1}}\fi
\expandafter\ifx\csname urlprefix\endcsname\relax\def\urlprefix{URL }\fi
\providecommand{\eprint}[2][]{\url{#2}}

\bibitem{donnelly1991}
Donnelly R~J 1991 {\em {Quantized vortices in helium II}\/} vol~2 (Cambridge
  University Press)

\bibitem{degnnes1966}
de~Gennes P~G 2018 {\em {Superconductivity of metals and alloys}\/} (CRC press)

\bibitem{matthews1999}
Matthews M~R, Anderson B~P, Haljan P~C, Hall D~S, Wieman C~E and Cornell E~A
  1999 {\em Phys. Rev. Lett.\/} {\bf 83} 2498

\bibitem{fetter2001}
Fetter A~L and Svidzinsky A~A 2001 {\em J. Phys. Condens. Matter\/} {\bf 13}
  R135

\bibitem{lundh1998}
Lundh E, Pethick C~J and Smith H 1998 {\em Phys. Rev. A\/} {\bf 58} 4816

\bibitem{dalfovo2000}
Dalfovo F and Modugno M 2000 {\em Phys. Rev. A\/} {\bf 61} 023605

\bibitem{seo2014}
Seo S~W, Choi J and Shin Y 2014 {\em Journal of the Korean Physical Society\/}
  {\bf 64} 53

\bibitem{srinivasan2006}
Srinivasan R 2006 {\em Pramana\/} {\bf 66} 3

\bibitem{madison2000}
Madison K~W, Chevy F, Wohlleben W and Dalibard J 2000 {\em Phys. Rev. Lett.\/}
  {\bf 84}(5) 806

\bibitem{abo-shaeer2001}
Abo-Shaeer J~R, Raman C, Vogels J~M and Ketterle W 2001 {\em Science\/} {\bf
  292} 476

\bibitem{hodby2001}
Hodby E, Hechenblaikner G, Hopkins S~A, Marag\`o O~M and Foot C~J 2001 {\em
  Phys. Rev. Lett.\/} {\bf 88}(1) 010405

\bibitem{engels2002}
Engels P, Coddington I, Haljan P~C and Cornell E~A 2002 {\em Phys. Rev.
  Lett.\/} {\bf 89}(10) 100403

\bibitem{inouye2001}
Inouye S, Gupta S, Rosenband T, Chikkatur A~P, G\"orlitz A, Gustavson T~L,
  Leanhardt A~E, Pritchard D~E and Ketterle W 2001 {\em Phys. Rev. Lett.\/}
  {\bf 87}(8) 080402

\bibitem{neely2010}
Neely T~W, Samson E~C, Bradley A~S, Davis M~J and Anderson B~P 2010 {\em Phys.
  Rev. Lett.\/} {\bf 104}(16) 160401

\bibitem{kwon2015}
Kwon W~J, Moon G, Seo S~W and Shin Y 2015 {\em Phys. Rev. A\/} {\bf 91}(5)
  053615

\bibitem{kwon2016}
Kwon W~J, Kim J~H, Seo S~W and Shin Y 2016 {\em Phys. Rev. Lett.\/} {\bf
  117}(24) 245301

\bibitem{lim2022}
Lim Y, Lee Y, Goo J, Bae D and Shin Y 2022 {\em New J. Phys.\/} {\bf 24} 083020

\bibitem{hadzibabic2006}
Hadzibabic Z, Kr{\"u}ger P, Cheneau M, Battelier B and Dalibard J 2006 {\em
  Nature\/} {\bf 441} 1118

\bibitem{schweikhard2007}
Schweikhard V, Tung S and Cornell E~A 2007 {\em Phys. Rev. Lett.\/} {\bf 99}(3)
  030401

\bibitem{choi2013}
Choi J, Seo S~W and Shin Y 2013 {\em Phys. Rev. Lett.\/} {\bf 110} 175302

\bibitem{seo2017}
Seo S~W, Ko B, Kim J~H and Shin Y 2017 {\em Scientific reports\/} {\bf 7} 4587

\bibitem{tsatsos2016}
Tsatsos M~C, Tavares P~E~S, Cidrim A, Fritsch A~R, Caracanhas M~A, dos Santos
  F~E~A, Barenghi C~F and Bagnato V~S 2016 {\em Phys. Rep.\/} {\bf 622} 1

\bibitem{henn2009}
Henn E~A~L, Seman J~A, Roati G, Magalhães K~M~F and Bagnato V~S 2009 {\em
  Phys. Rev. Lett.\/} {\bf 103} 045301

\bibitem{neely2013}
Neely T~W, Bradley A~S, Samson E~C, Rooney S~J, Wright E~M, Law K~J~H,
  Carretero-González R, Kevrekidis P~G, Davis M~J and Anderson B~P 2013 {\em
  Phys. Rev. Lett.\/} {\bf 111} 235301

\bibitem{kwon2014}
Kwon W~J, Moon G, Choi J, Seo S~W and Shin Y 2014 {\em Phys. Rev. A\/} {\bf
  90}(6) 063627

\bibitem{johnstone2019}
Johnstone S~P, Groszek A~J, Starkey P~T, Billington C~J, Simula T~P and
  Helmerson K 2019 {\em Science\/} {\bf 364} 1267

\bibitem{gauthier2019}
Gauthier G, Reeves M~T, Yu X, Bradley A~S, Baker M~A, Bell T~A,
  Rubinsztein-Dunlop H, Davis M~J and Neely T~W 2019 {\em Science\/} {\bf 364}
  1264

\bibitem{weiler2008}
Weiler C~N, Neely T~W, Scherer D~R, Bradley A~S, Davis M~J and Anderson B~P
  2008 {\em Nature\/} {\bf 455} 948

\bibitem{ko2019}
Ko B, Park J~W and Shin Y 2019 {\em Nat. Phys.\/} {\bf 15} 1227

\bibitem{goo2021}
Goo J, Lim Y and Shin Y 2021 {\em Phys. Rev. Lett.\/} {\bf 127}(11) 115701

\bibitem{goo2022}
Goo J, Lee Y, Lim Y, Bae D, Rabga T and Shin Y 2022 {\em Phys. Rev. Lett.\/}
  {\bf 128} 135701

\bibitem{rakonjac2016}
Rakonjac A, Marchant A~L, Billam T~P, Helm J~L, Yu M~M~H, Gardiner S~A and
  Cornish S~L 2016 {\em Phys. Rev. A\/} {\bf 93} 013607

\bibitem{wigley2016}
Wigley P~B, Everitt P~J, van~den Hengel A, Bastian J~W, Sooriyabandara M~A,
  McDonald G~D, Hardman K~S, Quinlivan C~D, Manju P, Kuhn C~C~N, Petersen I~R,
  Luiten A~N, Hope J~J, Robins N~P and Hush M~R 2016 {\em Sci Rep\/} {\bf 6}
  25890

\bibitem{barker2020}
Barker A~J, Style H, Luksch K, Sunami S, Garrick D, Hill F, Foot C~J and
  Bentine E 2020 {\em Mach. Learn.: Sci. Technol.\/} {\bf 1} 015007

\bibitem{ness2020}
Ness G, Vainbaum A, Shkedrov C, Florshaim Y and Sagi Y 2020 {\em Phys. Rev.
  Appl.\/} {\bf 14} 014011

\bibitem{davletov2020}
Davletov E~T, Tsyganok V~V, Khlebnikov V~A, Pershin D~A, Shaykin D~V and Akimov
  A~V 2020 {\em Phys. Rev. A\/} {\bf 102} 011302

\bibitem{rem2019}
Rem B~S, Käming N, Tarnowski M, Asteria L, Fläschner N, Becker C, Sengstock K
  and Weitenberg C 2019 {\em Nat. Phys.\/} {\bf 15} 917

\bibitem{bohrdt2019}
Bohrdt A, Chiu C~S, Ji G, Xu M, Greif D, Greiner M, Demler E, Grusdt F and Knap
  M 2019 {\em Nat. Phys.\/} {\bf 15} 921

\bibitem{kaming2021}
Käming N, Dawid A, Kottmann K, Lewenstein M, Sengstock K, Dauphin A and
  Weitenberg C 2021 {\em Mach. Learn.: Sci. Technol.\/} {\bf 2} 035037

\bibitem{hofer2021}
Hofer L~R, Krstajić M, Juhász P, Marchant A~L and Smith R~P 2021 {\em Mach.
  Learn.: Sci. Technol.\/} {\bf 2} 045008

\bibitem{guo2021}
Guo S, Fritsch A~R, Greenberg C, Spielman I~B and Zwolak J~P 2021 {\em Mach.
  Learn.: Sci. Technol.\/} {\bf 2} 035020

\bibitem{lode2021}
Lode A~U~J, Lin R, Büttner M, Papariello L, Lévêque C, Chitra R, Tsatsos
  M~C, Jaksch D and Molignini P 2021 {\em Phys. Rev. A\/} {\bf 104} L041301

\bibitem{sharma2022}
Sharma R and Simula T~P 2022 {\em Phys. Rev. A\/} {\bf 105} 033301

\bibitem{guo2022}
Guo S, Koh S~M, Fritsch A~R, Spielman I~B and Zwolak J~P 2022 {\em Phys. Rev.
  Res.\/} {\bf 4} 023163

\bibitem{metz2021}
Metz F, Polo J, Weber N and Busch T 2021 {\em Mach. Learn.: Sci. Technol.\/}
  {\bf 2} 035019

\bibitem{nikolenko2021}
Nikolenko S~I 2021 {\em {Synthetic Data for Deep Learning}\/} vol 174
  (Springer) ISBN 978-3-030-75177-7

\bibitem{demelo2022}
de~Melo C~M, Torralba A, Guibas L, DiCarlo J, Chellappa R and Hodgins J 2022
  {\em Trends in cognitive sciences\/} {\bf 26} 174

\bibitem{lu2023}
Lu Y, Wang H and Wei W 2023 {Machine Learning for Synthetic Data Generation: A
  Review} arXiv:2302.04062

\bibitem{kibble1976}
Kibble T~W~B 1976 {\em J. Phys. A\/} {\bf 9} 1387

\bibitem{zurek1985}
Zurek W~H 1985 {\em Nature\/} {\bf 317} 505

\bibitem{lim2021}
Lim Y, Goo J, Kwak H and Shin Y 2021 {\em Phys. Rev. A\/} {\bf 103} 063319

\bibitem{muruganandam2009}
Muruganandam P and Adhikari S~K 2009 {\em Comput. Phys. Commun.\/} {\bf 180}
  1888

\bibitem{vudragovic2012}
Vudragovi{\'c} D, Vidanovi{\'c} I, Bala{\v{z}} A, Muruganandam P and Adhikari
  S~K 2012 {\em Comput. Phys. Commun.\/} {\bf 183} 2021

\bibitem{esteve2006}
Esteve J, Trebbia J~B, Schumm T, Aspect A, Westbrook C~I and Bouchoule I 2006
  {\em Phys. Rev. Lett.\/} {\bf 96} 130403

\bibitem{sanner2010}
Sanner C, Su E~J, Keshet A, Gommers R, Shin Y, Huang W and Ketterle W 2010 {\em
  Phys. Rev. Lett.\/} {\bf 105} 040402

\bibitem{gomez2020}
Gómez-Ruiz F~J, Mayo J~J and del Campo A 2020 {\em Phys. Rev. Lett.\/} {\bf
  124} 240602

\bibitem{liu2021}
Liu X~P, Yao X~C, Deng Y, Wang X~Q, Wang Y~X, Huang C~J, Li X, Chen Y~A and Pan
  J~W 2021 {\em Phys. Rev. Lett.\/} {\bf 126} 185302

\bibitem{kim2022}
Kim M, Rabga T, Lee Y, Goo J, Bae D and Shin Y 2022 {\em Phys. Rev. A\/} {\bf
  106} L061301

\bibitem{delcampo2022}
del Campo A, Gómez-Ruiz F~J and Zhang H~Q 2022 {\em Phys. Rev. B\/} {\bf 106}
  L140101

\bibitem{rabga2023}
Rabga T, Lee Y, Bae D, Kim M and Shin Y 2023 {\em Phys. Rev. A\/} {\bf 108}(2)
  023315

\end{thebibliography}

\end{document}